\documentclass[]{IEEEtran}
\IEEEoverridecommandlockouts
\input epsf
\usepackage{graphicx}
\usepackage{url}
\usepackage{xcolor}
\usepackage{upgreek}

\usepackage{amsmath}
\usepackage[colorlinks=true,linkcolor=blue,urlcolor=blue,citecolor=blue]{hyperref}
\usepackage[colorinlistoftodos]{todonotes}

\begin{document}

\IEEEpubid{\makebox[\columnwidth]{979-8-3315-4683-0/26/\$31.00~\copyright2026 IEEE \hfill} \hspace{\columnsep}\makebox[\columnwidth]{ }}

\title{\LARGE An Optical System for Monitoring Coil Parasitic Motion and Mass Position for Tsinghua Tabletop Kibble Balance}

\author{Weibo Liu, Kang Ma, Nanjia Li, Wei Zhao, Songling Huang, Shisong Li$^\dagger$\\
Department of Electrical Engineering, Tsinghua University, Beijing 100084, China\\
$^\dagger$Email: shisongli@tsinghua.edu.cn}

\maketitle

\IEEEpubidadjcol

\begin{abstract}
This paper presents a novel seven-channel optical measurement system for monitoring coil parasitic motion and mass position in the Tsinghua Tabletop Kibble balance. The system employs seven spectrally-confocal displacement sensors arranged in a distributed configuration to simultaneously measure the coil's translational ($x_{\rm{c}}, y_{\rm{c}}$), rotational ($\theta_x,\theta_y$) degrees of freedom, and the mass position offset ($x_{\rm{m}}, y_{\rm{m}}$) due to corner errors. Three vertically oriented sensors target an equilateral triangle target rigidly connected to the coil, enabling real-time calculation of tilt angles through geometric relationships. Two horizontally oriented sensors measure the translational displacement of a frame target on the coil assembly. Two additional horizontal sensors monitor the mass position to quantify corner errors. The initial experimental setup has been completed, featuring sufficient resolution and minimal signal loss, providing a new approach for alignment adjustment and corner error compensation in high-precision Kibble balances.
\end{abstract}



\pagenumbering{gobble}

\section{Introduction}

The redefinition of the kilogram in the new SI based on the Planck constant has established the Kibble balance~\cite{Robinson_2016watt} as a pivotal instrument for realizing the mass unit, the kilogram. Recent researches have focused on developing compact, tabletop versions in conventional laboratory settings, thereby facilitating the distribution of new mass standards. The Tsinghua Tabletop Kibble balance~\cite{THUdesign2022} represents an ongoing endeavor in this direction.

Parasitic motions of the coil, including lateral translations ($x_{\rm{c}}, y_{\rm{c}}$) and tilts ($\theta_x,\theta_y$) relative to the magnetic field axis, introduce systematic errors in both voltage and force measurements. Existing monitoring methods are constrained in compact implementations. Conventional systems, which typically employ multi-interferometer setups or position-sensitive detectors (PSDs)~\cite{Fang_2020BIPM,NISTInvitedArticle2016}, require complex optical arrangements. These configurations are sensitive to misalignment, susceptible to signal loss, and less robust to integrate into the limited space of tabletop systems. Also, in the Tsinghua tabletop Kibble balance, the mass is loaded or unloaded in a mass pan above the pivot point, and the non-repeatable positioning of mass artifacts during loading, the corner error, can affect weighing results. Currently, monitoring of mass positions ($x_{\rm{m}}, y_{\rm{m}}$) typically relies on crude mechanical stops, lacking high-precision quantitative monitoring methods.

This paper addresses these challenges by presenting an innovative optical monitoring system based on spectrally-confocal displacement sensors. The system's compact, distributed design enables simultaneous, non-contact measurement of the coil parasitic motion and two mass position coordinates. 
\IEEEpubidadjcol

\section{System Design}
\label{sec:system_design}

\begin{figure}[htbp]
\centering
\includegraphics[width=0.85\linewidth]{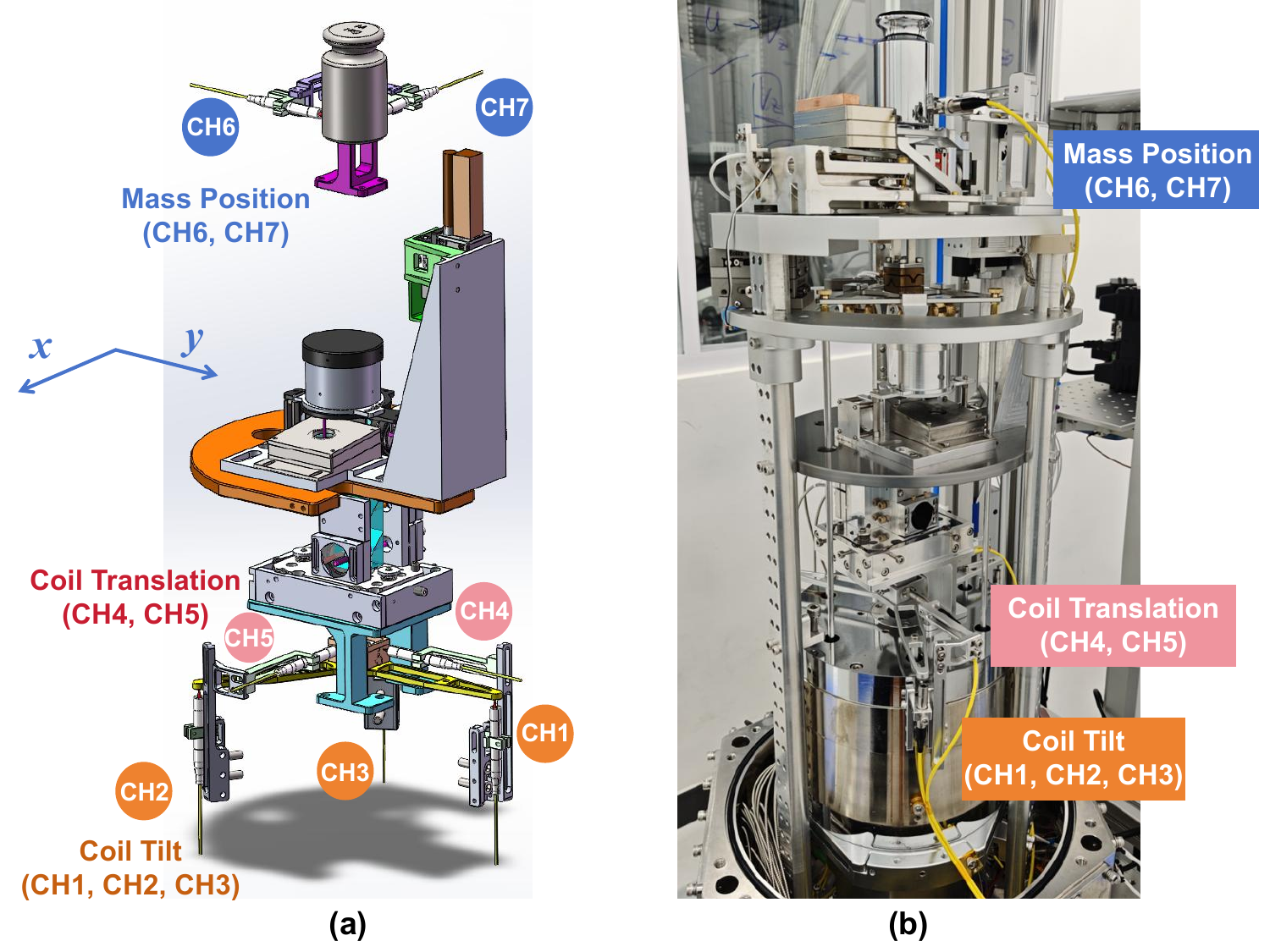}
\caption{(a) Schematic layout of the seven-channel monitoring system integrated with the THUKB coil and mass assembly. (b) Experimental setup.}
\label{fig:system_layout}
\end{figure}

The monitoring system is designed in Fig.~\ref{fig:system_layout}(a) as a peripheral module to the THUKB, operating without interference to the core electromagnetic measurements. Seven spectrally-confocal displacement sensors, with $\pm$1.5\,mm measurement range, and 0.1\,$\upmu$m repeatability, are employed. The sensors operate on the principle of axial chromatic dispersion, where a broadband light source is spectrally decomposed along the optical axis, and the wavelength of the light focused onto the target surface is detected to determine displacement with high precision. The sensor arrangement comprises: Sensors CH1, CH2, CH3 are for $\theta_x$, $\theta_y$ measurement. They are oriented vertically upward, targeting vertices A, B, C of the triangular target, which is rigidly connected to the coil frame via suspension rods. Sensors CH4, CH5 are for $x_{\rm{c}}, y_{\rm{c}}$ measurement, oriented horizontally, targeting orthogonal surfaces on the frame target. The frame target houses the corner cube assembly used for coil velocity measurement. It is secured to the triangular target, which is connected to the coil. Sensors CH6 and CH7 are horizontally oriented for monitoring the test mass position ($x_{\rm{m}}, y_{\rm{m}}$), and are aimed at orthogonal surfaces on the mass.    

\section{Measurement Principles and Models}
\label{sec:measurement_principles}

\subsection{Coil Tilt Monitoring}
The geometric relationship between vertex height changes measured by sensors $\Delta z$ and coil tilt angles $\theta$ is derived from rigid body kinematics. For an equilateral triangle target with arm length $L$=125\,mm, vertex angles $\phi_A = 0^\circ$, $\phi_B = 120^\circ$, $\phi_C = 240^\circ$, it yields
\begin{align}
\Delta z_A &= L(\theta_x \cos\phi_A + \theta_y \sin\phi_A), \\
\Delta z_B &= L(\theta_x \cos\phi_B + \theta_y \sin\phi_B), \\
\Delta z_C &= L(\theta_x \cos\phi_C + \theta_y \sin\phi_C).
\end{align}
Solving for the coil tilt $\theta_x$, $\theta_y$ yields:
\begin{equation}
\begin{bmatrix}
\theta_x \\ \theta_y
\end{bmatrix}
= \frac{1}{L}
\begin{bmatrix}
\frac{2}{3} & -\frac{1}{3} & -\frac{1}{3} \\
0 & \frac{1}{\sqrt{3}} & -\frac{1}{\sqrt{3}}
\end{bmatrix}
\begin{bmatrix}
\Delta z_A \\ \Delta z_B \\ \Delta z_C
\end{bmatrix}.
\end{equation}

This formulation inherently decouples tilt from translation, as uniform translation produces equal $\Delta z$ values that cancel in the tilt calculation. Given the sensors' repeatability of 0.1\,$\upmu$m for $\Delta z$, the repeatability for $\theta$ can reach 0.8\,$\upmu$rad.

\subsection{Coil Translation Monitoring}
For small displacements and near-orthogonal sensor alignment, the measured displacements $\Delta d_4$ and $\Delta d_5$ approximate the coil's translational components:
\begin{equation}
\Delta y_{\text{c}} \approx -\Delta d_4 / \cos(\alpha_4), \quad \Delta x_{\text{c}} \approx -\Delta d_5 / \cos(\alpha_5).
\end{equation}
where $\alpha_4$, $\alpha_5$ are small installation misalignment angles. With proper alignment ($\alpha \approx 0$), $\Delta y_{\text{c}} \approx -\Delta d_4$ and $\Delta x_{\text{c}} \approx -\Delta d_5$.

\subsection{Mass Position Monitoring}
Due to the near-orthogonal relationship, the mass position offsets $\Delta x_{\text{m}}$ and $\Delta y_{\text{m}}$ (horizontal displacement of the center of mass of the weight) are obtained:
\begin{align}
\Delta y_{\text{m}} =\Delta d_6+R_{\rm{m}}\sqrt{1-\frac{\Delta x_{\text{m}}^2}{R_{\rm{m}}^2}}-R_{\rm{m}},
\\
\Delta x_{\text{m}} =\Delta d_7+R_{\rm{m}}\sqrt{1-\frac{\Delta y_{\text{m}}^2}{R_{\rm{m}}^2}}-R_{\rm{m}}.
\end{align}
In principle, the above is a nonlinear system of simultaneous equations requiring joint solution. When $|\Delta y_{\text{m}}|$ and $|\Delta x_{\text{m}}|$ are sufficiently small compared to $R_{\text{m}}$ (Radius of the mass), an approximate solution can be obtained from the measured displacements $\Delta d_6$, $\Delta d_7$ by sensors CH6 and CH7.

\begin{equation}
\Delta y_{\text{m}} \approx \Delta d_6, \quad \Delta x_{\text{m}} \approx \Delta d_7.
\end{equation}

For the THUKB weighing unit, an initial corner error adjustment was performed using a 100\,g weight on an extended 110\,mm × 64\,mm weighing platform to achieve consistency at the milligram level. In order to achieve a measurement uncertainty of 10\,$\upmu$g, the lateral displacement of the 1\,kg mass on the pan must be constrained to less than 0.1\,mm during the weighing process. The sensors' precision is sufficient to meet this requirement.

\section{Experimental Implementation}
\label{sec:experimental}

The monitoring system was integrated into the THUKB experimental setup in Fig.~\ref{fig:system_layout}(b). Sensors were mounted on rigid brackets with fine-adjustment mechanisms for precise targeting. The triangular and frame targets were machined from aluminum. To enhance the reflectivity of the triangular target and the frame target, flat reflectors with a diameter of 10\,mm were affixed at the optical beam alignment points. All sensors are connected to a multi-channel data acquisition system, and the host PC software enables real-time data processing, visualisation, and storage capabilities.

\begin{figure}[htbp]
\centering
\includegraphics[width=1.05\linewidth]{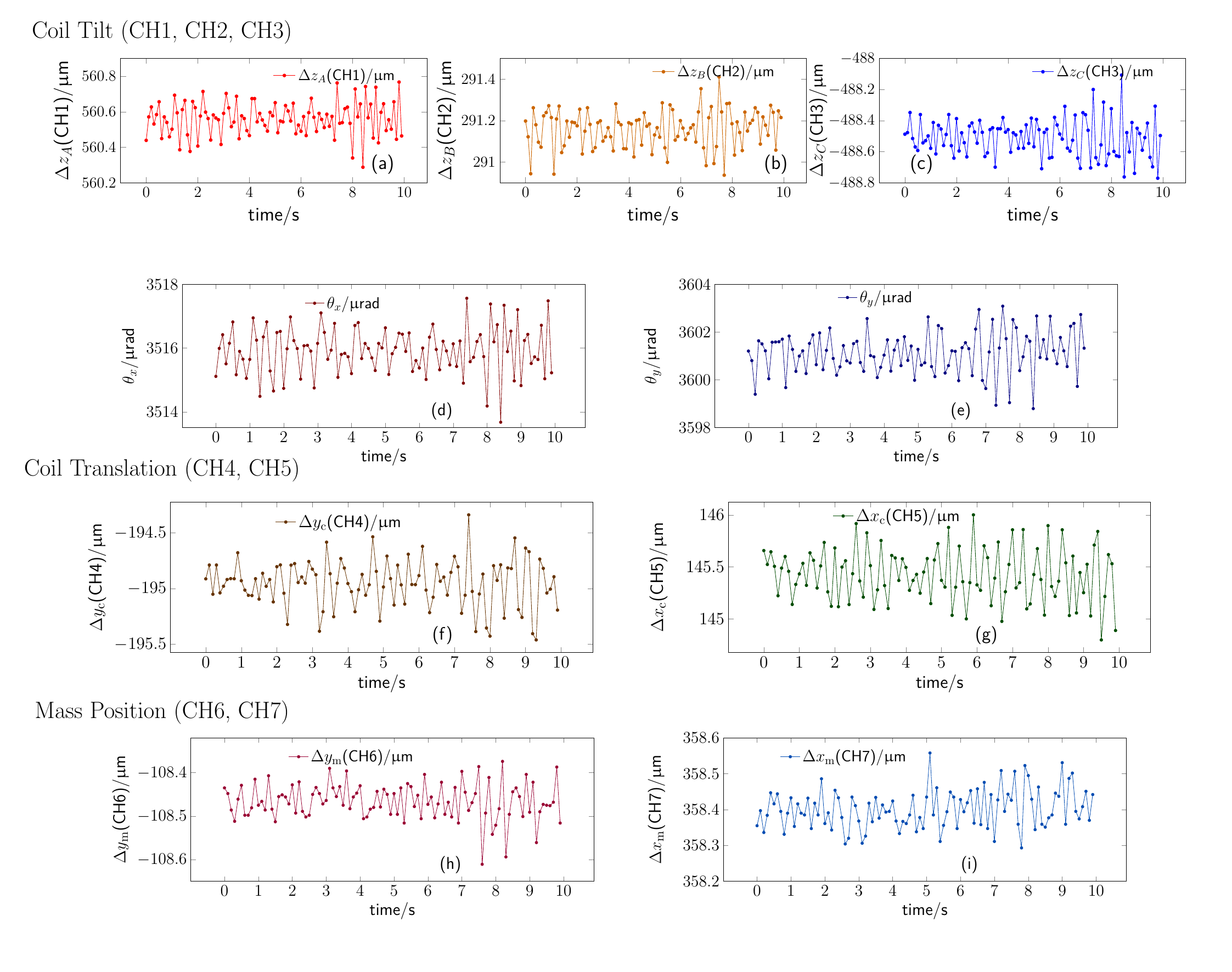}
\caption{Preliminary test signals from the seven sensors indicate that the sensors are capable of normal readings and operation.}
\label{fig:EXPresult}
\end{figure}

The system demonstrated excellent robustness during operation, with all channels maintaining signals as in Fig~\ref{fig:EXPresult}. Manual adjustments to the coil suspension and XY stage produced measurable and correct responses in the channels.

\vspace{-0.3cm}

\vspace{-0.3cm}
\section{Conclusion}
\label{sec:conclusion}
This paper proposes and implements an innovative optical monitoring system based on a seven-channel spectral confocal displacement sensor, tailored to the specific requirements of THUKB. The design methodology is also applicable to other precision instrumentation requiring multi-degree-of-freedom motion monitoring.

\section*{Acknowledgment}
This work was supported by the National Natural Science Foundation of China under Grant 52377011.


\end{document}